\documentclass[aps,prl,twocolumn,eqsecnum,superscriptaddress,floatfix,10pt]{revtex4-2}
\usepackage{amsmath,amssymb,amsfonts,stmaryrd,wasysym,graphicx,multirow,color,textcomp,subfigure,nicefrac,float,enumitem}
\usepackage{url}
\usepackage[colorlinks=true,citecolor=blue,urlcolor=blue,linkcolor=blue] {hyperref}

\usepackage{amsmath}

\usepackage{qcircuit}
\usepackage{braket}
\usepackage{romannum}

\usepackage{color,soul} 
\usepackage{bm}
\usepackage[normalem]{ulem}

\def\be{\begin{equation}}
\def\ee{\end{equation}}
\def\bea{\begin{eqnarray}}
\def\eea{\end{eqnarray}}

\newcommand{\cyan}[1]{{\color{cyan}{\it {#1}}}} 

\counterwithout{equation}{section}

\begin{document}

\title{Floquet Chiral Quantum Walk in Quantum Computer}
\author{Chan Bin Bark}
\address{Department of Physics, Hanyang University, Seoul 04763, Republic of Korea}
\author{Youngseok Kim}
\email{youngseok.kim1@ibm.com}
\address{IBM Quantum, IBM T.J. Watson Research Center, Yorktown Heights, NY, USA}
\author{Moon Jip Park}
\email{moonjippark@hanyang.ac.kr}
\address{Department of Physics, Hanyang University, Seoul 04763, Republic of Korea}
\date{\today}

\begin{abstract}
Chiral edge states in quantum Hall effect are the paradigmatic example of the quasi-particle with chirality. In even space-time dimensions, the Nielsen-Ninomiya theorem strictly forbids the chiral states in physical isolation. The exceptions to this theorem only occur in the presence of non-locality\cite{PhysRevD.16.3031,PhysRevX.13.011003}, non-Hermiticity\cite{PhysRevLett.127.196404}, or by embedding the system at the boundary of the higher-dimensional bulk\cite{RevModPhys.82.3045}. In this work, using the IBM quantum computer platform, we realize the floquet chiral quantum walk enabled by non-locality. The unitary time evolution operator is described by the effective floquet Hamiltonian with infinitely long-ranged coupling. We find that the chiral wave packets lack the common features of the conventional wave phenomena such as Anderson localization. The absence of localization is witnessed by the robustness against the external perturbations. However, the intrinsic quantum errors of the current quantum device give rise to the finite lifetime where the chiral wave packet eventually disperses in the long-time limit. Nevertheless, we observe the stability of the chiral wave by comparing it with the conventional non-chiral model.
\end{abstract}
\maketitle

\cyan{Introduction--} Nielsen-Ninomiya no-go theorem\cite{NIELSEN1981219,NIELSEN198120,NIELSEN1981173}, applicable in any even dimensional lattice systems, states that the number of the chiral fermion species always appear as pairs. That is, no lattice systems are allowed to have net-chirality if the Hamiltonian is (i) local, (ii) Hermitian, and (iii) translational symmetric. Cracking any of the above conditions gives rise to the path to realize the chiral fermions isolated from its pair\cite{PhysRevLett.116.133903, KAPLAN1992342,PhysRevLett.123.190403,PhysRevB.102.041119,PELISSETTO1988177,SHAMIR1994590,Li2015,Semenoff_2012,PhysRevB.86.115133,Chen2023}. The representative examples that bypass the theorem are the classes of topological insulators. In the quantum Hall insulators\cite{RevModPhys.95.011002}, the isolated chiral edge states (say right-moving mode) are realized on its boundary while its anti-chiral partner (left-moving mode) is spatially separated in the other edge\cite{PhysRevB.89.085101}. Chiral fermions, once they are realized in isolated systems, are of particular interest due to the possible applications based on unusual phenomena such as chiral anomaly and resistivity against localization\cite{kharzeev2014chiral,burkov2015chiral,parameswaran2014probing}.

\begin{figure}[t!]
    \centering
    \includegraphics[width=\linewidth]{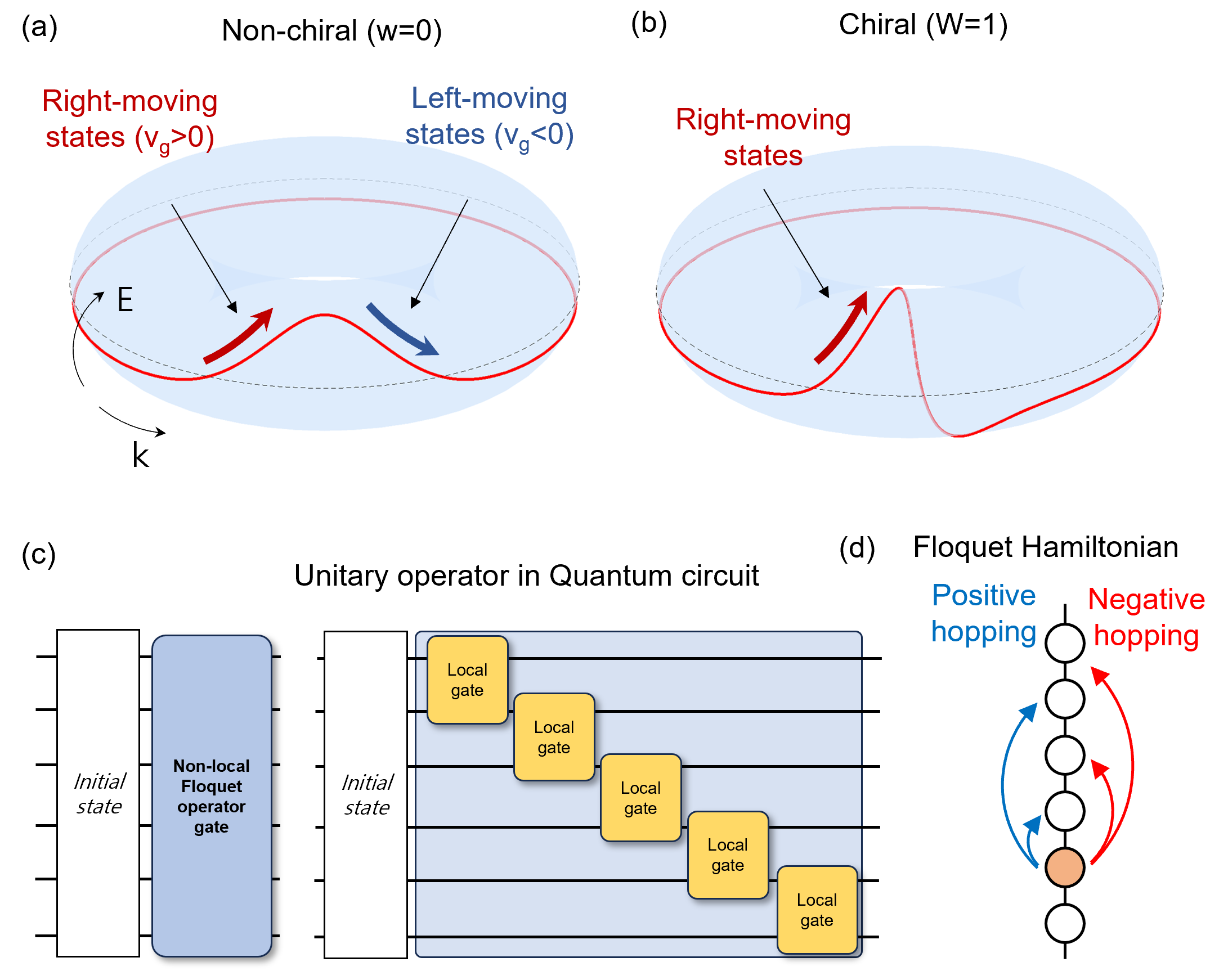}  
    \caption{(a)-(b) Chiral and non-Chiral Band structure of the floquet evolution operator in the quasi-energy-momentum plane. (a) In the case of the non-chiral state, the chiral and anti-chiral states appear as pairs. (b) However, in the chiral state, the winding of the band structure along the quasi-energy direction results in non-zero net chirality. (c) Illustration of non-local unitary operator consists of local unitary gates. Subsequent application of the local operation generates the non-local Floquet circuit. (d) Corresponding Hamiltonian in the real space lattice. Long-ranged coupling with alternating signs produces destructive interference, which results in chiral wave propagation.
}
    \label{fig1}
\end{figure}

Meanwhile, the advent of noisy intermediate-scale quantum (NISQ) devices provides a toolbox with the ability to design the desired Hamiltonian of complex quantum systems\cite{feynman2018simulating,deutsch1985quantum,boghosian1998simulating,georgescu2014quantum,das2008colloquium,cloet2019opportunities,kassal2011simulating,daley2022practical,whitfield2011simulation,cervera2018exact,smith2019simulating,e12112268,Preskill2018quantumcomputingin,Kim2023}. In this work, we realize the floquet chiral wave functions enabled by non-local couplings using the current IBM Quantum processor. The non-local time-evolution operator, $\hat{U}_\textrm{NL}^\textrm{F}$, is constructed by the products of the successive local quantum operations as [See Fig. \ref{fig1} (c)]:
\bea
\hat{U}_\textrm{NL}^\textrm{F}=\prod_{\langle i,j \rangle } \hat{U}_{ij}^L,
\eea
where $i,j$ indicates the qubit sites of the nearest neighbor. The unitary operator is described by the floquet band theory, where the quasi-eigenenergy of the floquet operator is periodic, $\epsilon(k)\in [0,2\pi]$, as well as the quasi-momentum $k\in [0,2\pi]$. The periodicity of the floquet quasienergy allows to have topologically non-trivial band structure that has the net-winding along the quasi-energy direction [See Fig. \ref{fig1}(b)]. The winding number manifests as the isolated right-moving chiral mode with the positive group velocity, $v_g=\partial \epsilon /\partial k >0$, without the left-moving pair.

\begin{figure*}[t!]
\centering
\includegraphics[width=\linewidth]{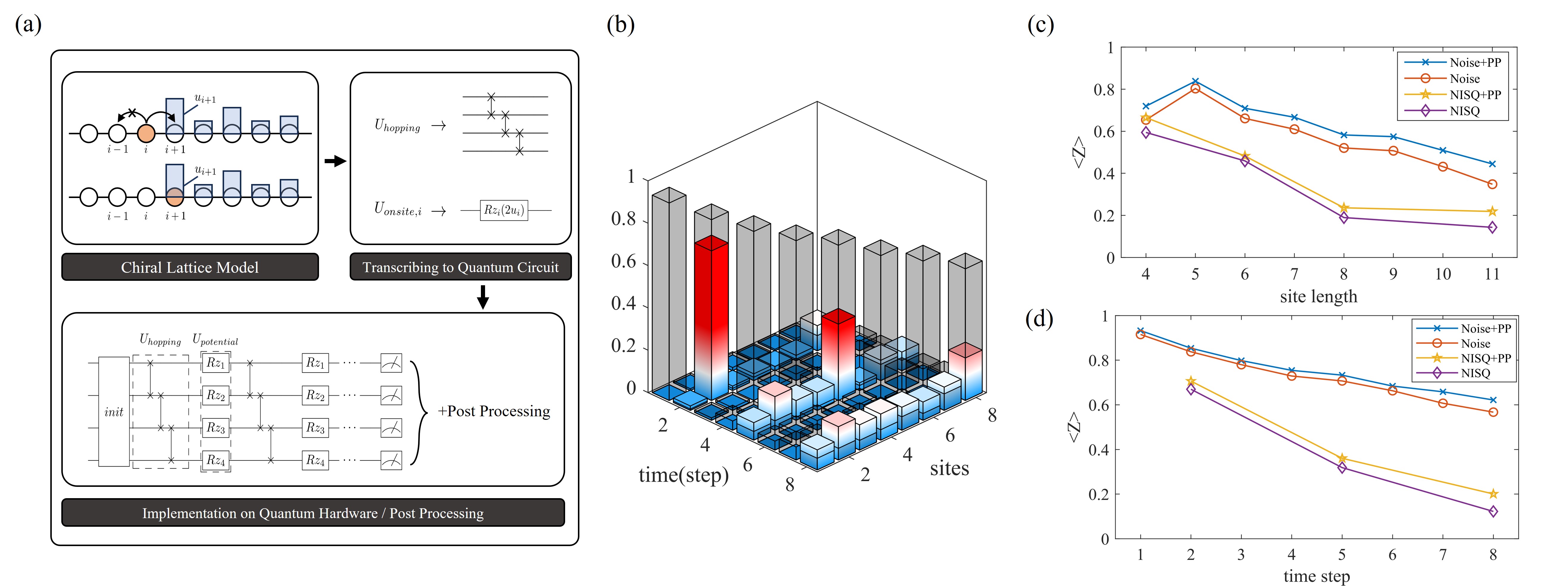}

\caption{(a) Schematic illustration of the quantum circuit and the particle moving in the one-dimensional periodic lattice. The FCQW is composed of the products of the two unitary operators, $U_{\textrm{hopping}}$ and $U_{\textrm{onsite}}$. We transcribe it in quantum circuit languages to implement in NISQ devices(\textit{ibm\_hanoi}). (b) Propagation of chiral wave function in $L=8$ lattice without the external potential. The gray bars represent the corresponding results from noise simulations using Qiskit fake provider class($FakeHanoi$), while the colored bars correspond to the outcomes from the real device experiment on \textit{ibm\_hanoi}. In real device experiment, total 7000 shot is executed and we apply dynamical decoupling to mitigate error rates. We observe the evident chiral wave propagation. The discrepancy between the real device and the noisy simulator is caused by the intrinsic noise. (c)-(d) the comparison of the wave-packet peak amplitude as a function of (c) the total system size and (d) time steps respectively. PP represents the post-processing. While the ideal value of the peak amplitude is exactly 1, the intrinsic error in the noisy simulator and the real device causes the reduction of the amplitude. The post-processing described in Eq. \eqref{pp} shows recognizable improvement (c) as a function of the time step $t=L$, so the error rate rapidly escalates in proportion to $O(L^2)$. (d) when the site length is fixed at a $L=8$, the error rate shows a linear increase which is proportional to $O(L)$.}
    \label{fig2}
\end{figure*}

In this work, we have conducted quantum simulations of the floquet chiral quantum walk(FCQW) on IBM Quantum quantum devices, encoded in quantum circuit language. The simulations of FCQW showcase the unidirectional behavior of wave functions in isolated one-dimensional systems. The destructive interference is shown in the real-space Hamiltonian where the long-range coupling blocks the hopping of the quasiparticle in one of the directions [Fig. \ref{fig1}(d)]. The realized chiral quantum walk exhibits inherent topological immunity against backscattering. To validate the robustness, we introduced external potential barriers and observed the wave function propagations, thereby reconfirming their stability against external perturbations. Furthermore, we contrast the robustness of the chiral states by observing the Anderson localization in the conventional non-chiral quantum model in the continuous time domain.

\begin{figure}[t!]
\centering
\includegraphics[width=1\linewidth]{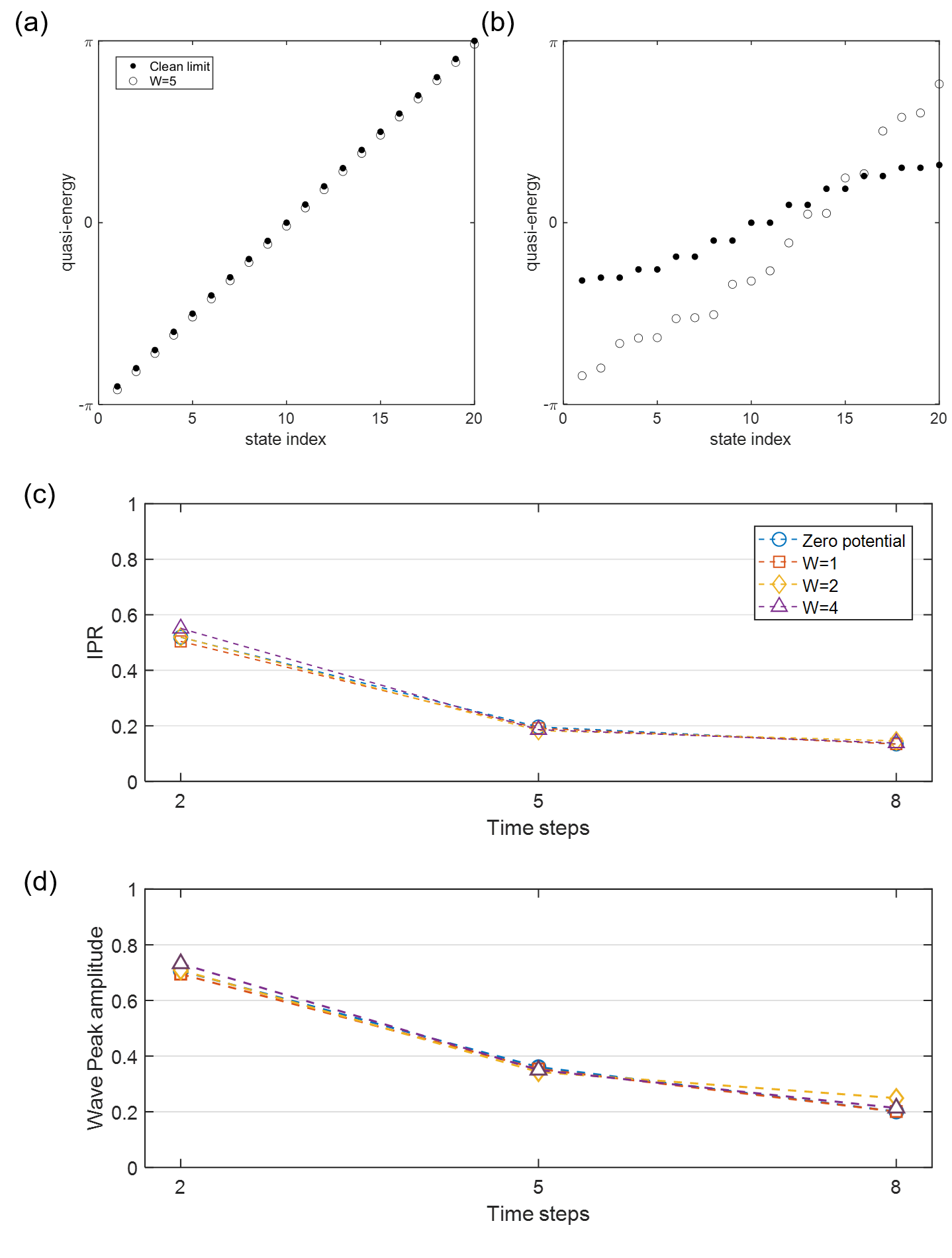}

\caption{(a)-(b) Numerical simulation of the quasi-energy spectra in the presence of the disorder in (a) chiral (b) non-chiral wave functions. In the presence of the strong random potential, the non-chiral states show drastic changes in the eigenstates distributions, signifying the transitions to localized states. In contrast, the spectra of the chiral wave functions only show the overall shift of the energy.  (c)-(d) Experimental comparison of (c) IPR and (d) wave-packet peak amplitude as a function of the potential strength. Experimental condition is same with the case of $W=0$ (Fig \ref{fig2}(b)). Even in the strongest potential strength $W=4\sim 2\pi$, the IPR and the wave-packet peak amplitude show a relative difference of less than 10$\%$ with that of the clean limit ($W=0$).}
    \label{fig3}
\end{figure}

\begin{figure*}[t!]
\centering
\includegraphics[width=0.9\linewidth]{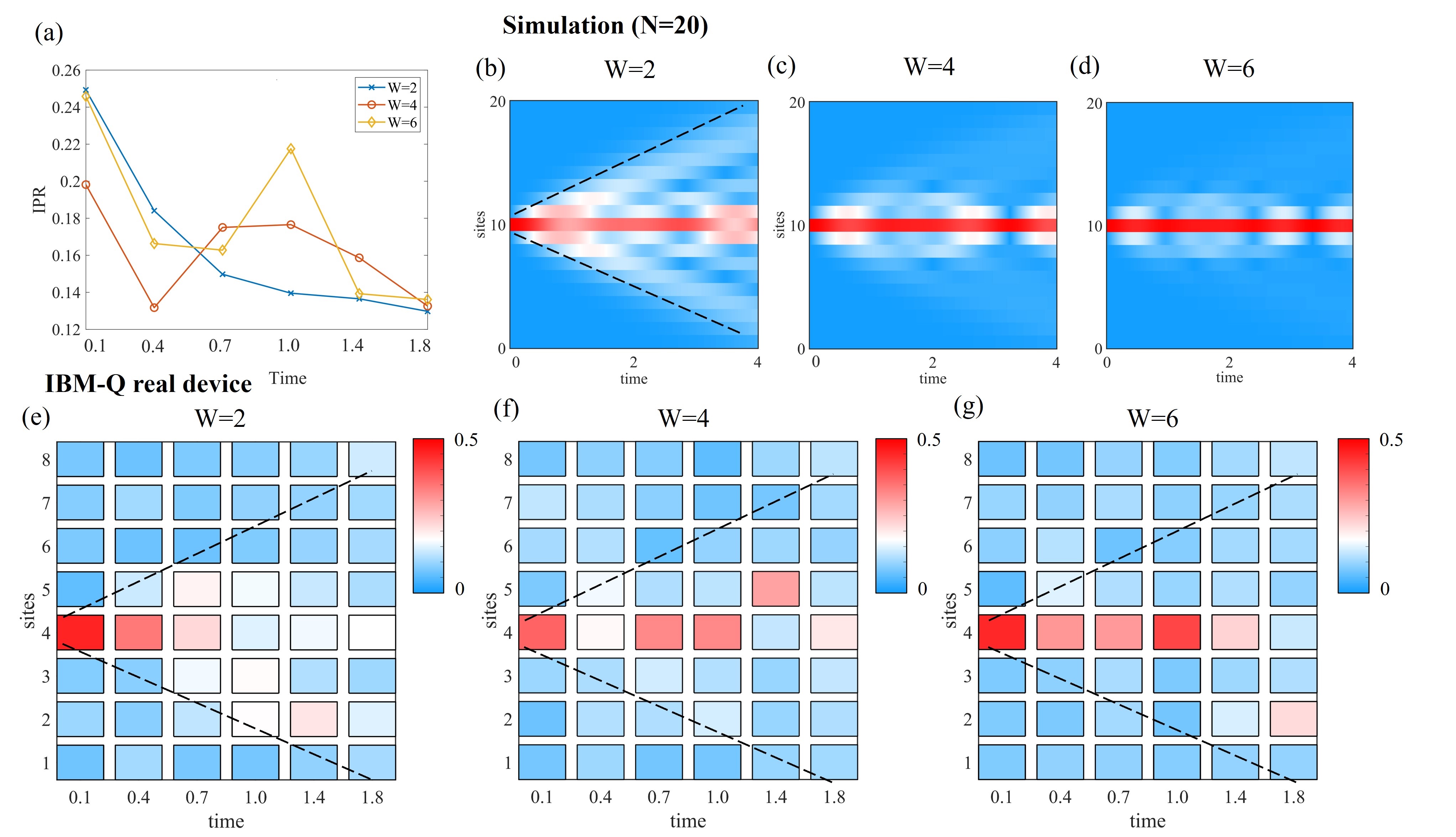}

\caption{(a) Time evolution of the IPR of non-chiral wave function as a function of the potential strength. (b)-(d) numerical simulation ($L=20$) and (e)-(g) experimentally observed wave propagation ($L=8$, \textit{ibmq\_kolkata}) of the non-chiral wave functions. The real device measurement is performed with a total 7000 execution times per data point and we apply dynamical decoupling to mitigate error rates. In the strong potential barrier, the propagation is strongly suppressed due to the confinement of the wave packet. Black-dotted lines are the reference to the dispersion of the wave packet without potential. In both simulation and experimental results, we find the fragility of the non-chiral lattice model against Anderson localization. }
    \label{fig4}
\end{figure*}

\cyan{Floquet chiral quantum walk--} We represent the periodic lattice in $L$-qubit system model by mapping $|0\rangle$ and $|1\rangle$ qubit states as unoccupied and occupied states respectively. The single discrete time-step evolution, $|\psi (t+1)\rangle=\hat{U}_{\textrm{NL}}^F|\psi (t)\rangle$, is described by the floquet unitary time-evolution operator $\hat{U}_{\textrm{NL}}^F={\hat{U}}_{\textrm{hopping}}{\hat{U}}_{\textrm{onsite}}$, comprised of the kinetic term (${\hat{U}}_{\textrm{hopping}}$) and the local potential terms (${\hat{U}}_{\textrm{onsite}}$) that accounts for the backscattering. The kinetic term plays a role in uni-directional hopping as, 
\bea
{\hat{U}}_{\textrm{hopping}}=\sum_{i=1}^L |i+1\rangle \langle i | ,
\label{Eq:2}
\eea
where $|i\rangle$ indicates the localized states at $i$-th qubit, and we assume the periodic boundary $|L+1 \rangle =|1\rangle$. The corresponding effective floquet Hamiltonian $H_F=-i \ln U^F_\textrm{NL}$ has power law decaying non-local couplings in the presence of the net-chirality.

The chiral propagation of the FCQW is the manifestation of the non-trivial floquet topological invariant\cite{PhysRevB.96.155118}. Unlike the continuous time domain, the discrete quantum walk is described by the quasienergy eigenvalues $\epsilon_n(k)\in [0,2\pi]$, where the eigenstate $|n(k)\rangle$ satisfies $\hat{U}_{\textrm{NL}}^F |n(k)\rangle=e^{i\epsilon_n(k)}|n(k)\rangle$, where $k$ is the quasi-momentum. In general, the quasienergy-quasimomentum space forms the two-dimensional torus $T^2$, where we can consider the winding of the quasi-energy band along the quasi-energy direction. For instance, the dispersion of \eqref{Eq:2} is given as, $\epsilon(k)=k$, that has non-zero chirality [See Fig. \ref{fig1}(b)].

Formally, under the non-trivial homotopy group, $\pi_1(\textrm{U}(N))=\mathbb{Z}$ (A class in the Altland-Zirnbauer classification), we can define the winding number ($w\in \mathbb{Z}$) of the quasi-energy band, which is given as, 
\bea
w=\frac{1}{2\pi i}\oint dk [U^F_k]^\dagger \partial_k U^F_k,
\eea
where $U_k^F$ is the Fourier transformed time-evolution operator in the momentum space. The non-zero winding number indicates $w$ net-chirality (difference between the number of the chiral and anti-chiral modes.) of the systems. We point out that the isolated chiral wave can only be manifested in the floquet system due to the periodic nature of the quasi-energy. For instance, in the continuous time domain, the chiral motion in Eq. \eqref{Eq:2} corresponds to a limit of the Hatano-Nelson model, which inevitably introduces the non-Hermiticity in the Hamiltonian.

\cyan{Transcribing chiral lattice model into quantum circuit--} To implement the FCQW on a real quantum circuit device, we utilize the successive application of the local SWAP operator on the nearest neighbor qubits. The hopping term is expressed as a product of the SWAP operators acting on the nearest neighbor qubits, where the hopping term is explicitly given as,
\begin{gather}
{
\hat{U}_{hopping} = \prod_{<i,j>}^N (SWAP)_{ij}
}.
\end{gather}
The implementation of the periodic boundary condition does not need the swap operators between the long-distanced two end sites. For instance, when $L=4$, the SWAP operator allows the exchange of quantum states between adjacent lattice sites. 
Accordingly, $\hat{U}_{\textrm{hopping}}$  can be represented in a quantum circuit, as follows,
\begin{equation}
\hat{U}_{hopping} = \begin{pmatrix}
0&1&0&0 \\
0&0&1&0 \\
0&0&0&1 \\
1&0&0&0
    \end{pmatrix}
=
    \begin{tabular}{c}
        \vspace{-.1em}\\
        \Qcircuit @C=1.4em @R=1.1em {
        &\qw & \qswap & \qw& \qw & \qw  \\ 
        &\qw & \qswap \qwx & \qswap & \qw & \qw \\
        &\qw & \qw & \qswap \qwx& \qswap & \qw\\ 
        &\qw & \qw & \qw & \qswap \qwx & \qw 
        }
        \vspace{2em}\hspace{0.5em}\\
    \end{tabular}
\end{equation}

Using \textit{ibm\_hanoi} processor, we designed the quantum circuit that emulates FCQW in $L=8$ qubit lattice  [See Fig. \ref{fig2}(a) and the supplementary material for the detailed implementation]. The wave-propagation is measured at the discrete time steps $(t=2,5,8)$ where the quantum circuit is evolved with the time evolution of the interval ($0\le t \le 8$). The initial state is prepared to be localized at a single site ($i=1$). The results are measured with 7000 shots and compared against the results from noisy simulator. We find the evident chiral propagation of the wave packet [colored bars in Fig \ref{fig2}(a)] that agrees well with the results from noisy-simulator [gray bars in Fig \ref{fig2}(a)]. In addition, during the time evolution, we observe the continuous dispersion of the localized wave packet, evidenced by lowered wave function amplitude at the packet center with the enhanced background noise. 

This backscattering-like noise would not be seen in the ideal circuit. By analyzing the measured data, we find that the intrinsic noise of the processor is the dominant source of the dispersion. In our mapping, for example, the undesired bit-flip corresponds to the particle creation and annihilation that maps to the sectors with different particle numbers. The noise induced dispersion is globally observed without the signature of specific spatial correlation. In the transcribed circuit, the SWAP gate is composed of three C-NOT gates. As such, the time evolution operator contains $(N-1)$ SWAP gates per step. The rate of the random bit-flip increases in proportion to both non-local depth (number of CNOT gates) and the increase in the system size. To mitigate the dispersion in the particle number space, we integrate the many-body sectors with different particle numbers rather than discarding them. Considering the uncorrelated random nature of the noise, we can count the normalized probability density as,
\bea
\textrm{P}_i(t) = \langle \psi (t)|\hat{N}_i|\psi(t)\rangle / (\sum_{j=1}^{N}
\langle \psi (t)|\hat{N}_j|\psi(t)\rangle ),
\label{pp}
\eea
where $\hat{N}_i$ is the particle number operator at $i$-th site. Fig.\ref{fig2} (c)-(d) compares the measured data with and without the error mitigation. We observe recognizable improvement in the error rate observed in both real devices and noisy simulators. 

\cyan{Topological robustness against disorder--} We experimentally validate the robustness against external potential in the coexistence with the intrinsic quantum error. 

The onsite potential $\hat{U}_{\textrm{onsite}}$ can be designed by performing the transcribed RZ gates to each site,
$\hat{U}_{\textrm{onsite}} = \prod_{i=1}^L RZ_{i}(2W u_i)$, where $W$ is the controllable potential strength. $u_i$ determines the shape of the potential. The RZ gate is a one-qubit gate that applies a phase to a qubit.
\begin{equation} \notag
{
RZ_{i}(\theta) = \begin{pmatrix}
e^{-i \frac{\theta}{2}} &0\\
0&e^{+i\frac{\theta}{2}}
    \end{pmatrix} 
= \begin{tabular}{c}
\vspace{-1em}\\
\Qcircuit @C=0.9em @R=1em {
& \qw & \gate{RZ_{i}(\theta)} & \qw 
}
\vspace{0.2em}\hspace{1em}\\
\end{tabular}
},
\end{equation}
 where the additional phase is gained under RZ gate if the corresponding site is occupied. To examine the wave confinement, we impose the square box potential barrier next to the initially localized site such that $u_{i=2,3,7,8}=1$ otherwise $u_i=0$. The strength of the potential barrier is varied up to $W=4$, which is comparable to the maximal window of the quasi-energy in the floquet systems. Exceeding more than $W=2\pi$ results in the reduction of the potential due to the periodic nature of the floquet quasi-energy.
 
 Fig. \ref{fig3}(c),(d) show the comparison of the inverse-participation ratio (IPR) and the peak amplitude of the wave packet as a function of the strength of the potential barrier. IPR which is defined by $(IPR)=\sum_i\left| \psi_i \right|^4$ can conveniently describe localized(extended) behaviors of quantum systems\cite{DF9705000055,Wegner1980InversePR}. For various ranges of the potential barrier strength, we do not find the signature of the reduction in the chiral propagation. The measured difference of the IPR and the peak amplitude between the clean limit and the strong potential $W=4$ ranges up to $10\%$, which is ignorably smaller compared to the dispersion effect induced by the intrinsic quantum errors. 
 
The robustness against the external potential is theoretically examined by calculating the change in the quasi-energy spectra. Fig. \ref{fig3}(a),(b) exemplifies the comparison of the quasi-energy spectra in the presence of the random Anderson disorder with chiral ($w=1$) and non-chiral ($w=0$) respectively. The introduction of the disorder immediately causes the change in the level statistics of the non-chiral quasi-eigenvalue spectra, indicating the Anderson localization transition. On the other hand, the level statistics of the chiral states remain unchanged, and the overall shift of the energy is only observed. This feature is rather unusual even when compared to the chiral edge state of the two-dimensional Chern insulator. In general, the reduction in the transmission is observed in the edge transport of the two-dimensional Chern insulator. This is because the localized wave packets are generally linear superpositions of both bulk and edge states. Strong disorder comparable to the bulk gap causes a small finite hybridization between the bulk and edge. However, in the case of the isolated chiral states, the bulk contribution in the transmission is conceptually absent, which shows the extreme robustness compared to the chiral edge states in the topological insulators.

\cyan{Comparison with non-chiral wave function--}
To contrast the behavior of the chiral and non-chiral states, we show the Anderson localization in the non-chiral states\cite{PhysRevLett.42.673,kappus1981anomaly}. We realize the one-dimensional tight-binding chain using the trotterized Hamiltonian, which is given as,
\begin{gather}
\hat{H} = \sum_{i=1}^L -[\hat{\sigma}_{i}^{x} \hat{\sigma}_{i+1}^{x} + \hat{\sigma}_{i}^{y} \hat{\sigma}_{i+1}^{y}] + Wu_{i} \hat{\sigma}_{i}^{z}
,
\end{gather}
where $\hat{\sigma}^{\mu}_{i}$ is $\mu$-th Pauli operator at site $i$. We simulate the continuous evolution of the above Hamiltonian using the trotterized circuit of \textit{ibmq\_kolkata} device. (See supplementary material for the detailed circuit implementation of the non-chiral states). Upon the time evolution, we choose the initial wave function localized at the site $i=4$, where the measurement is performed during the six-time intervals from $t=0.1$ to $t=2.0$. Fig. \ref{fig4} (e)-(g) shows the comparison of the wave propagation while varying the potential strength [The corresponding numerical simulation is shown in Fig. \ref{fig4} (b)-(d)]. Fig. \ref{fig4} shows the time evolution of the IPR. Unlike the chiral wave function, the IPR of the non-chiral wave fluctuates strongly as a function of the potential strength. Furthermore, the long-time enhancement of the IPR indicates the strong localization feature driven by the potential barrier. In the clean limit ($W=0$), the wave propagation occurs in both directions, although asymmetry in the propagation is observed due to the intrinsic noise. As the potential strength increases up to $W=6$, the wave function confinement appears, and it survives long-time up to $t=1.8$ in both results from simulation and \textit{ibmq\_kolkata} device. The confinement is the direct evidence of the fragility against Anderson localization. This feature is in direct contrast to the FCQW. 

\cyan{Discussions--}  Our proposal of the chiral states has no counterpart in the continuous time domain since the topological origin of the winding number is based on the floquet structure of the discrete-time evolution operator. Realization of the chiral wave function can be extended in higher-spatial dimensions. For instance, in two dimensions, the fermion doubling forbids the odd number of the two-dimensional Dirac fermion without breaking time-reversal symmetry\cite{PhysRevD.26.468,RevModPhys.55.775}. Recently the realization of the single Dirac cone has been also proposed using the quantum optics toolbox\cite{PhysRevX.13.011003}. In addition, our proposal is fully capable of utilizing many-body wave functions. Studying the chiral wave function in the presence of the many-body wave function would open a new avenue to study the intriguing many-body phenomena such as the fractionalized quasi-particles.

\bibliography{reference}

\begin{thebibliography}{47}%
\makeatletter
\providecommand \@ifxundefined [1]{%
 \@ifx{#1\undefined}
}%
\providecommand \@ifnum [1]{%
 \ifnum #1\expandafter \@firstoftwo
 \else \expandafter \@secondoftwo
 \fi
}%
\providecommand \@ifx [1]{%
 \ifx #1\expandafter \@firstoftwo
 \else \expandafter \@secondoftwo
 \fi
}%
\providecommand \natexlab [1]{#1}%
\providecommand \enquote  [1]{``#1''}%
\providecommand \bibnamefont  [1]{#1}%
\providecommand \bibfnamefont [1]{#1}%
\providecommand \citenamefont [1]{#1}%
\providecommand \href@noop [0]{\@secondoftwo}%
\providecommand \href [0]{\begingroup \@sanitize@url \@href}%
\providecommand \@href[1]{\@@startlink{#1}\@@href}%
\providecommand \@@href[1]{\endgroup#1\@@endlink}%
\providecommand \@sanitize@url [0]{\catcode `\\12\catcode `\$12\catcode `\&12\catcode `\#12\catcode `\^12\catcode `\_12\catcode `\%12\relax}%
\providecommand \@@startlink[1]{}%
\providecommand \@@endlink[0]{}%
\providecommand \url  [0]{\begingroup\@sanitize@url \@url }%
\providecommand \@url [1]{\endgroup\@href {#1}{\urlprefix }}%
\providecommand \urlprefix  [0]{URL }%
\providecommand \Eprint [0]{\href }%
\providecommand \doibase [0]{https://doi.org/}%
\providecommand \selectlanguage [0]{\@gobble}%
\providecommand \bibinfo  [0]{\@secondoftwo}%
\providecommand \bibfield  [0]{\@secondoftwo}%
\providecommand \translation [1]{[#1]}%
\providecommand \BibitemOpen [0]{}%
\providecommand \bibitemStop [0]{}%
\providecommand \bibitemNoStop [0]{.\EOS\space}%
\providecommand \EOS [0]{\spacefactor3000\relax}%
\providecommand \BibitemShut  [1]{\csname bibitem#1\endcsname}%
\let\auto@bib@innerbib\@empty
\bibitem [{\citenamefont {Susskind}(1977)}]{PhysRevD.16.3031}%
  \BibitemOpen
  \bibfield  {author} {\bibinfo {author} {\bibfnamefont {L.}~\bibnamefont {Susskind}},\ }\bibfield  {title} {\bibinfo {title} {Lattice fermions},\ }\href {https://doi.org/10.1103/PhysRevD.16.3031} {\bibfield  {journal} {\bibinfo  {journal} {Phys. Rev. D}\ }\textbf {\bibinfo {volume} {16}},\ \bibinfo {pages} {3031} (\bibinfo {year} {1977})}\BibitemShut {NoStop}%
\bibitem [{\citenamefont {Kim}\ \emph {et~al.}(2023{\natexlab{a}})\citenamefont {Kim}, \citenamefont {Bagrets}, \citenamefont {Micklitz},\ and\ \citenamefont {Altland}}]{PhysRevX.13.011003}%
  \BibitemOpen
  \bibfield  {author} {\bibinfo {author} {\bibfnamefont {K.~W.}\ \bibnamefont {Kim}}, \bibinfo {author} {\bibfnamefont {D.}~\bibnamefont {Bagrets}}, \bibinfo {author} {\bibfnamefont {T.}~\bibnamefont {Micklitz}},\ and\ \bibinfo {author} {\bibfnamefont {A.}~\bibnamefont {Altland}},\ }\bibfield  {title} {\bibinfo {title} {Floquet simulators for topological surface states in isolation},\ }\href {https://doi.org/10.1103/PhysRevX.13.011003} {\bibfield  {journal} {\bibinfo  {journal} {Phys. Rev. X}\ }\textbf {\bibinfo {volume} {13}},\ \bibinfo {pages} {011003} (\bibinfo {year} {2023}{\natexlab{a}})}\BibitemShut {NoStop}%
\bibitem [{\citenamefont {Bessho}\ and\ \citenamefont {Sato}(2021)}]{PhysRevLett.127.196404}%
  \BibitemOpen
  \bibfield  {author} {\bibinfo {author} {\bibfnamefont {T.}~\bibnamefont {Bessho}}\ and\ \bibinfo {author} {\bibfnamefont {M.}~\bibnamefont {Sato}},\ }\bibfield  {title} {\bibinfo {title} {Nielsen-ninomiya theorem with bulk topology: Duality in floquet and non-hermitian systems},\ }\href {https://doi.org/10.1103/PhysRevLett.127.196404} {\bibfield  {journal} {\bibinfo  {journal} {Phys. Rev. Lett.}\ }\textbf {\bibinfo {volume} {127}},\ \bibinfo {pages} {196404} (\bibinfo {year} {2021})}\BibitemShut {NoStop}%
\bibitem [{\citenamefont {Hasan}\ and\ \citenamefont {Kane}(2010)}]{RevModPhys.82.3045}%
  \BibitemOpen
  \bibfield  {author} {\bibinfo {author} {\bibfnamefont {M.~Z.}\ \bibnamefont {Hasan}}\ and\ \bibinfo {author} {\bibfnamefont {C.~L.}\ \bibnamefont {Kane}},\ }\bibfield  {title} {\bibinfo {title} {Colloquium: Topological insulators},\ }\href {https://doi.org/10.1103/RevModPhys.82.3045} {\bibfield  {journal} {\bibinfo  {journal} {Rev. Mod. Phys.}\ }\textbf {\bibinfo {volume} {82}},\ \bibinfo {pages} {3045} (\bibinfo {year} {2010})}\BibitemShut {NoStop}%
\bibitem [{\citenamefont {Nielsen}\ and\ \citenamefont {Ninomiya}(1981{\natexlab{a}})}]{NIELSEN1981219}%
  \BibitemOpen
  \bibfield  {author} {\bibinfo {author} {\bibfnamefont {H.}~\bibnamefont {Nielsen}}\ and\ \bibinfo {author} {\bibfnamefont {M.}~\bibnamefont {Ninomiya}},\ }\bibfield  {title} {\bibinfo {title} {A no-go theorem for regularizing chiral fermions},\ }\href {https://doi.org/https://doi.org/10.1016/0370-2693(81)91026-1} {\bibfield  {journal} {\bibinfo  {journal} {Physics Letters B}\ }\textbf {\bibinfo {volume} {105}},\ \bibinfo {pages} {219} (\bibinfo {year} {1981}{\natexlab{a}})}\BibitemShut {NoStop}%
\bibitem [{\citenamefont {Nielsen}\ and\ \citenamefont {Ninomiya}(1981{\natexlab{b}})}]{NIELSEN198120}%
  \BibitemOpen
  \bibfield  {author} {\bibinfo {author} {\bibfnamefont {H.}~\bibnamefont {Nielsen}}\ and\ \bibinfo {author} {\bibfnamefont {M.}~\bibnamefont {Ninomiya}},\ }\bibfield  {title} {\bibinfo {title} {Absence of neutrinos on a lattice: (i). proof by homotopy theory},\ }\href {https://doi.org/https://doi.org/10.1016/0550-3213(81)90361-8} {\bibfield  {journal} {\bibinfo  {journal} {Nuclear Physics B}\ }\textbf {\bibinfo {volume} {185}},\ \bibinfo {pages} {20} (\bibinfo {year} {1981}{\natexlab{b}})}\BibitemShut {NoStop}%
\bibitem [{\citenamefont {Nielsen}\ and\ \citenamefont {Ninomiya}(1981{\natexlab{c}})}]{NIELSEN1981173}%
  \BibitemOpen
  \bibfield  {author} {\bibinfo {author} {\bibfnamefont {H.}~\bibnamefont {Nielsen}}\ and\ \bibinfo {author} {\bibfnamefont {M.}~\bibnamefont {Ninomiya}},\ }\bibfield  {title} {\bibinfo {title} {Absence of neutrinos on a lattice: (ii). intuitive topological proof},\ }\href {https://doi.org/https://doi.org/10.1016/0550-3213(81)90524-1} {\bibfield  {journal} {\bibinfo  {journal} {Nuclear Physics B}\ }\textbf {\bibinfo {volume} {193}},\ \bibinfo {pages} {173} (\bibinfo {year} {1981}{\natexlab{c}})}\BibitemShut {NoStop}%
\bibitem [{\citenamefont {Lee}(2016)}]{PhysRevLett.116.133903}%
  \BibitemOpen
  \bibfield  {author} {\bibinfo {author} {\bibfnamefont {T.~E.}\ \bibnamefont {Lee}},\ }\bibfield  {title} {\bibinfo {title} {Anomalous edge state in a non-hermitian lattice},\ }\href {https://doi.org/10.1103/PhysRevLett.116.133903} {\bibfield  {journal} {\bibinfo  {journal} {Phys. Rev. Lett.}\ }\textbf {\bibinfo {volume} {116}},\ \bibinfo {pages} {133903} (\bibinfo {year} {2016})}\BibitemShut {NoStop}%
\bibitem [{\citenamefont {Kaplan}(1992)}]{KAPLAN1992342}%
  \BibitemOpen
  \bibfield  {author} {\bibinfo {author} {\bibfnamefont {D.~B.}\ \bibnamefont {Kaplan}},\ }\bibfield  {title} {\bibinfo {title} {A method for simulating chiral fermions on the lattice},\ }\href {https://doi.org/https://doi.org/10.1016/0370-2693(92)91112-M} {\bibfield  {journal} {\bibinfo  {journal} {Physics Letters B}\ }\textbf {\bibinfo {volume} {288}},\ \bibinfo {pages} {342} (\bibinfo {year} {1992})}\BibitemShut {NoStop}%
\bibitem [{\citenamefont {H\"ockendorf}\ \emph {et~al.}(2019)\citenamefont {H\"ockendorf}, \citenamefont {Alvermann},\ and\ \citenamefont {Fehske}}]{PhysRevLett.123.190403}%
  \BibitemOpen
  \bibfield  {author} {\bibinfo {author} {\bibfnamefont {B.}~\bibnamefont {H\"ockendorf}}, \bibinfo {author} {\bibfnamefont {A.}~\bibnamefont {Alvermann}},\ and\ \bibinfo {author} {\bibfnamefont {H.}~\bibnamefont {Fehske}},\ }\bibfield  {title} {\bibinfo {title} {Non-hermitian boundary state engineering in anomalous floquet topological insulators},\ }\href {https://doi.org/10.1103/PhysRevLett.123.190403} {\bibfield  {journal} {\bibinfo  {journal} {Phys. Rev. Lett.}\ }\textbf {\bibinfo {volume} {123}},\ \bibinfo {pages} {190403} (\bibinfo {year} {2019})}\BibitemShut {NoStop}%
\bibitem [{\citenamefont {Wu}\ and\ \citenamefont {An}(2020)}]{PhysRevB.102.041119}%
  \BibitemOpen
  \bibfield  {author} {\bibinfo {author} {\bibfnamefont {H.}~\bibnamefont {Wu}}\ and\ \bibinfo {author} {\bibfnamefont {J.-H.}\ \bibnamefont {An}},\ }\bibfield  {title} {\bibinfo {title} {Floquet topological phases of non-hermitian systems},\ }\href {https://doi.org/10.1103/PhysRevB.102.041119} {\bibfield  {journal} {\bibinfo  {journal} {Phys. Rev. B}\ }\textbf {\bibinfo {volume} {102}},\ \bibinfo {pages} {041119} (\bibinfo {year} {2020})}\BibitemShut {NoStop}%
\bibitem [{\citenamefont {Pelissetto}(1988)}]{PELISSETTO1988177}%
  \BibitemOpen
  \bibfield  {author} {\bibinfo {author} {\bibfnamefont {A.}~\bibnamefont {Pelissetto}},\ }\bibfield  {title} {\bibinfo {title} {Lattice non-local chiral fermions},\ }\href {https://doi.org/https://doi.org/10.1016/0003-4916(88)90299-0} {\bibfield  {journal} {\bibinfo  {journal} {Annals of Physics}\ }\textbf {\bibinfo {volume} {182}},\ \bibinfo {pages} {177} (\bibinfo {year} {1988})}\BibitemShut {NoStop}%
\bibitem [{\citenamefont {Shamir}(1994)}]{SHAMIR1994590}%
  \BibitemOpen
  \bibfield  {author} {\bibinfo {author} {\bibfnamefont {Y.}~\bibnamefont {Shamir}},\ }\bibfield  {title} {\bibinfo {title} {Constraints on the existence of chiral fermions in interacting lattice theories},\ }\href {https://doi.org/https://doi.org/10.1016/0920-5632(94)90454-5} {\bibfield  {journal} {\bibinfo  {journal} {Nuclear Physics B - Proceedings Supplements}\ }\textbf {\bibinfo {volume} {34}},\ \bibinfo {pages} {590} (\bibinfo {year} {1994})},\ \bibinfo {note} {proceedings of the International Symposium on Lattice Field Theory}\BibitemShut {NoStop}%
\bibitem [{\citenamefont {Li}\ and\ \citenamefont {Sarma}(2015)}]{Li2015}%
  \BibitemOpen
  \bibfield  {author} {\bibinfo {author} {\bibfnamefont {X.}~\bibnamefont {Li}}\ and\ \bibinfo {author} {\bibfnamefont {S.~D.}\ \bibnamefont {Sarma}},\ }\bibfield  {title} {\bibinfo {title} {Exotic topological density waves in cold atomic rydberg-dressed fermions},\ }\href {https://doi.org/10.1038/ncomms8137} {\bibfield  {journal} {\bibinfo  {journal} {Nature Communications}\ }\textbf {\bibinfo {volume} {6}},\ \bibinfo {pages} {7137} (\bibinfo {year} {2015})}\BibitemShut {NoStop}%
\bibitem [{\citenamefont {Semenoff}(2012)}]{Semenoff_2012}%
  \BibitemOpen
  \bibfield  {author} {\bibinfo {author} {\bibfnamefont {G.~W.}\ \bibnamefont {Semenoff}},\ }\bibfield  {title} {\bibinfo {title} {Chiral symmetry breaking in graphene},\ }\href {https://doi.org/10.1088/0031-8949/2012/T146/014016} {\bibfield  {journal} {\bibinfo  {journal} {Physica Scripta}\ }\textbf {\bibinfo {volume} {2012}},\ \bibinfo {pages} {014016} (\bibinfo {year} {2012})}\BibitemShut {NoStop}%
\bibitem [{\citenamefont {Zyuzin}\ and\ \citenamefont {Burkov}(2012)}]{PhysRevB.86.115133}%
  \BibitemOpen
  \bibfield  {author} {\bibinfo {author} {\bibfnamefont {A.~A.}\ \bibnamefont {Zyuzin}}\ and\ \bibinfo {author} {\bibfnamefont {A.~A.}\ \bibnamefont {Burkov}},\ }\bibfield  {title} {\bibinfo {title} {Topological response in weyl semimetals and the chiral anomaly},\ }\href {https://doi.org/10.1103/PhysRevB.86.115133} {\bibfield  {journal} {\bibinfo  {journal} {Phys. Rev. B}\ }\textbf {\bibinfo {volume} {86}},\ \bibinfo {pages} {115133} (\bibinfo {year} {2012})}\BibitemShut {NoStop}%
\bibitem [{\citenamefont {Chen}\ \emph {et~al.}(2023)\citenamefont {Chen}, \citenamefont {Wu}, \citenamefont {Xi}, \citenamefont {Yi},\ and\ \citenamefont {Yue}}]{Chen2023}%
  \BibitemOpen
  \bibfield  {author} {\bibinfo {author} {\bibfnamefont {W.-Q.}\ \bibnamefont {Chen}}, \bibinfo {author} {\bibfnamefont {Y.-S.}\ \bibnamefont {Wu}}, \bibinfo {author} {\bibfnamefont {W.}~\bibnamefont {Xi}}, \bibinfo {author} {\bibfnamefont {W.-Z.}\ \bibnamefont {Yi}},\ and\ \bibinfo {author} {\bibfnamefont {G.}~\bibnamefont {Yue}},\ }\bibfield  {title} {\bibinfo {title} {Fate of quantum anomalies for 1d lattice chiral fermion with a simple non-hermitian hamiltonian},\ }\href {https://doi.org/10.1007/JHEP05(2023)090} {\bibfield  {journal} {\bibinfo  {journal} {Journal of High Energy Physics}\ }\textbf {\bibinfo {volume} {2023}},\ \bibinfo {pages} {90} (\bibinfo {year} {2023})}\BibitemShut {NoStop}%
\bibitem [{\citenamefont {Chang}\ \emph {et~al.}(2023)\citenamefont {Chang}, \citenamefont {Liu},\ and\ \citenamefont {MacDonald}}]{RevModPhys.95.011002}%
  \BibitemOpen
  \bibfield  {author} {\bibinfo {author} {\bibfnamefont {C.-Z.}\ \bibnamefont {Chang}}, \bibinfo {author} {\bibfnamefont {C.-X.}\ \bibnamefont {Liu}},\ and\ \bibinfo {author} {\bibfnamefont {A.~H.}\ \bibnamefont {MacDonald}},\ }\bibfield  {title} {\bibinfo {title} {Colloquium: Quantum anomalous hall effect},\ }\href {https://doi.org/10.1103/RevModPhys.95.011002} {\bibfield  {journal} {\bibinfo  {journal} {Rev. Mod. Phys.}\ }\textbf {\bibinfo {volume} {95}},\ \bibinfo {pages} {011002} (\bibinfo {year} {2023})}\BibitemShut {NoStop}%
\bibitem [{\citenamefont {Teo}\ and\ \citenamefont {Kane}(2014)}]{PhysRevB.89.085101}%
  \BibitemOpen
  \bibfield  {author} {\bibinfo {author} {\bibfnamefont {J.~C.~Y.}\ \bibnamefont {Teo}}\ and\ \bibinfo {author} {\bibfnamefont {C.~L.}\ \bibnamefont {Kane}},\ }\bibfield  {title} {\bibinfo {title} {From luttinger liquid to non-abelian quantum hall states},\ }\href {https://doi.org/10.1103/PhysRevB.89.085101} {\bibfield  {journal} {\bibinfo  {journal} {Phys. Rev. B}\ }\textbf {\bibinfo {volume} {89}},\ \bibinfo {pages} {085101} (\bibinfo {year} {2014})}\BibitemShut {NoStop}%
\bibitem [{\citenamefont {Kharzeev}(2014)}]{kharzeev2014chiral}%
  \BibitemOpen
  \bibfield  {author} {\bibinfo {author} {\bibfnamefont {D.~E.}\ \bibnamefont {Kharzeev}},\ }\bibfield  {title} {\bibinfo {title} {The chiral magnetic effect and anomaly-induced transport},\ }\href@noop {} {\bibfield  {journal} {\bibinfo  {journal} {Progress in Particle and Nuclear Physics}\ }\textbf {\bibinfo {volume} {75}},\ \bibinfo {pages} {133} (\bibinfo {year} {2014})}\BibitemShut {NoStop}%
\bibitem [{\citenamefont {Burkov}(2015)}]{burkov2015chiral}%
  \BibitemOpen
  \bibfield  {author} {\bibinfo {author} {\bibfnamefont {A.}~\bibnamefont {Burkov}},\ }\bibfield  {title} {\bibinfo {title} {Chiral anomaly and transport in weyl metals},\ }\href@noop {} {\bibfield  {journal} {\bibinfo  {journal} {Journal of Physics: Condensed Matter}\ }\textbf {\bibinfo {volume} {27}},\ \bibinfo {pages} {113201} (\bibinfo {year} {2015})}\BibitemShut {NoStop}%
\bibitem [{\citenamefont {Parameswaran}\ \emph {et~al.}(2014)\citenamefont {Parameswaran}, \citenamefont {Grover}, \citenamefont {Abanin}, \citenamefont {Pesin},\ and\ \citenamefont {Vishwanath}}]{parameswaran2014probing}%
  \BibitemOpen
  \bibfield  {author} {\bibinfo {author} {\bibfnamefont {S.}~\bibnamefont {Parameswaran}}, \bibinfo {author} {\bibfnamefont {T.}~\bibnamefont {Grover}}, \bibinfo {author} {\bibfnamefont {D.}~\bibnamefont {Abanin}}, \bibinfo {author} {\bibfnamefont {D.}~\bibnamefont {Pesin}},\ and\ \bibinfo {author} {\bibfnamefont {A.}~\bibnamefont {Vishwanath}},\ }\bibfield  {title} {\bibinfo {title} {Probing the chiral anomaly with nonlocal transport in three-dimensional topological semimetals},\ }\href@noop {} {\bibfield  {journal} {\bibinfo  {journal} {Physical Review X}\ }\textbf {\bibinfo {volume} {4}},\ \bibinfo {pages} {031035} (\bibinfo {year} {2014})}\BibitemShut {NoStop}%
\bibitem [{\citenamefont {Feynman}\ \emph {et~al.}(2018)\citenamefont {Feynman} \emph {et~al.}}]{feynman2018simulating}%
  \BibitemOpen
  \bibfield  {author} {\bibinfo {author} {\bibfnamefont {R.~P.}\ \bibnamefont {Feynman}} \emph {et~al.},\ }\bibfield  {title} {\bibinfo {title} {Simulating physics with computers},\ }\href@noop {} {\bibfield  {journal} {\bibinfo  {journal} {Int. j. Theor. phys}\ }\textbf {\bibinfo {volume} {21}} (\bibinfo {year} {2018})}\BibitemShut {NoStop}%
\bibitem [{\citenamefont {Deutsch}(1985)}]{deutsch1985quantum}%
  \BibitemOpen
  \bibfield  {author} {\bibinfo {author} {\bibfnamefont {D.}~\bibnamefont {Deutsch}},\ }\bibfield  {title} {\bibinfo {title} {Quantum theory, the church--turing principle and the universal quantum computer},\ }\href@noop {} {\bibfield  {journal} {\bibinfo  {journal} {Proceedings of the Royal Society of London. A. Mathematical and Physical Sciences}\ }\textbf {\bibinfo {volume} {400}},\ \bibinfo {pages} {97} (\bibinfo {year} {1985})}\BibitemShut {NoStop}%
\bibitem [{\citenamefont {Boghosian}\ and\ \citenamefont {Taylor~IV}(1998)}]{boghosian1998simulating}%
  \BibitemOpen
  \bibfield  {author} {\bibinfo {author} {\bibfnamefont {B.~M.}\ \bibnamefont {Boghosian}}\ and\ \bibinfo {author} {\bibfnamefont {W.}~\bibnamefont {Taylor~IV}},\ }\bibfield  {title} {\bibinfo {title} {Simulating quantum mechanics on a quantum computer},\ }\href@noop {} {\bibfield  {journal} {\bibinfo  {journal} {Physica D: Nonlinear Phenomena}\ }\textbf {\bibinfo {volume} {120}},\ \bibinfo {pages} {30} (\bibinfo {year} {1998})}\BibitemShut {NoStop}%
\bibitem [{\citenamefont {Georgescu}\ \emph {et~al.}(2014)\citenamefont {Georgescu}, \citenamefont {Ashhab},\ and\ \citenamefont {Nori}}]{georgescu2014quantum}%
  \BibitemOpen
  \bibfield  {author} {\bibinfo {author} {\bibfnamefont {I.~M.}\ \bibnamefont {Georgescu}}, \bibinfo {author} {\bibfnamefont {S.}~\bibnamefont {Ashhab}},\ and\ \bibinfo {author} {\bibfnamefont {F.}~\bibnamefont {Nori}},\ }\bibfield  {title} {\bibinfo {title} {Quantum simulation},\ }\href@noop {} {\bibfield  {journal} {\bibinfo  {journal} {Reviews of Modern Physics}\ }\textbf {\bibinfo {volume} {86}},\ \bibinfo {pages} {153} (\bibinfo {year} {2014})}\BibitemShut {NoStop}%
\bibitem [{\citenamefont {Das}\ and\ \citenamefont {Chakrabarti}(2008)}]{das2008colloquium}%
  \BibitemOpen
  \bibfield  {author} {\bibinfo {author} {\bibfnamefont {A.}~\bibnamefont {Das}}\ and\ \bibinfo {author} {\bibfnamefont {B.~K.}\ \bibnamefont {Chakrabarti}},\ }\bibfield  {title} {\bibinfo {title} {Colloquium: Quantum annealing and analog quantum computation},\ }\href@noop {} {\bibfield  {journal} {\bibinfo  {journal} {Reviews of Modern Physics}\ }\textbf {\bibinfo {volume} {80}},\ \bibinfo {pages} {1061} (\bibinfo {year} {2008})}\BibitemShut {NoStop}%
\bibitem [{\citenamefont {Clo{\"e}t}\ \emph {et~al.}(2019)\citenamefont {Clo{\"e}t}, \citenamefont {Dietrich}, \citenamefont {Arrington}, \citenamefont {Bazavov}, \citenamefont {Bishof}, \citenamefont {Freese}, \citenamefont {Gorshkov}, \citenamefont {Grassellino}, \citenamefont {Hafidi}, \citenamefont {Jacob} \emph {et~al.}}]{cloet2019opportunities}%
  \BibitemOpen
  \bibfield  {author} {\bibinfo {author} {\bibfnamefont {I.~C.}\ \bibnamefont {Clo{\"e}t}}, \bibinfo {author} {\bibfnamefont {M.~R.}\ \bibnamefont {Dietrich}}, \bibinfo {author} {\bibfnamefont {J.}~\bibnamefont {Arrington}}, \bibinfo {author} {\bibfnamefont {A.}~\bibnamefont {Bazavov}}, \bibinfo {author} {\bibfnamefont {M.}~\bibnamefont {Bishof}}, \bibinfo {author} {\bibfnamefont {A.}~\bibnamefont {Freese}}, \bibinfo {author} {\bibfnamefont {A.~V.}\ \bibnamefont {Gorshkov}}, \bibinfo {author} {\bibfnamefont {A.}~\bibnamefont {Grassellino}}, \bibinfo {author} {\bibfnamefont {K.}~\bibnamefont {Hafidi}}, \bibinfo {author} {\bibfnamefont {Z.}~\bibnamefont {Jacob}}, \emph {et~al.},\ }\bibfield  {title} {\bibinfo {title} {Opportunities for nuclear physics \& quantum information science},\ }\href@noop {} {\bibfield  {journal} {\bibinfo  {journal} {arXiv preprint arXiv:1903.05453}\ } (\bibinfo {year} {2019})}\BibitemShut {NoStop}%
\bibitem [{\citenamefont {Kassal}\ \emph {et~al.}(2011)\citenamefont {Kassal}, \citenamefont {Whitfield}, \citenamefont {Perdomo-Ortiz}, \citenamefont {Yung},\ and\ \citenamefont {Aspuru-Guzik}}]{kassal2011simulating}%
  \BibitemOpen
  \bibfield  {author} {\bibinfo {author} {\bibfnamefont {I.}~\bibnamefont {Kassal}}, \bibinfo {author} {\bibfnamefont {J.~D.}\ \bibnamefont {Whitfield}}, \bibinfo {author} {\bibfnamefont {A.}~\bibnamefont {Perdomo-Ortiz}}, \bibinfo {author} {\bibfnamefont {M.-H.}\ \bibnamefont {Yung}},\ and\ \bibinfo {author} {\bibfnamefont {A.}~\bibnamefont {Aspuru-Guzik}},\ }\bibfield  {title} {\bibinfo {title} {Simulating chemistry using quantum computers},\ }\href@noop {} {\bibfield  {journal} {\bibinfo  {journal} {Annual review of physical chemistry}\ }\textbf {\bibinfo {volume} {62}},\ \bibinfo {pages} {185} (\bibinfo {year} {2011})}\BibitemShut {NoStop}%
\bibitem [{\citenamefont {Daley}\ \emph {et~al.}(2022)\citenamefont {Daley}, \citenamefont {Bloch}, \citenamefont {Kokail}, \citenamefont {Flannigan}, \citenamefont {Pearson}, \citenamefont {Troyer},\ and\ \citenamefont {Zoller}}]{daley2022practical}%
  \BibitemOpen
  \bibfield  {author} {\bibinfo {author} {\bibfnamefont {A.~J.}\ \bibnamefont {Daley}}, \bibinfo {author} {\bibfnamefont {I.}~\bibnamefont {Bloch}}, \bibinfo {author} {\bibfnamefont {C.}~\bibnamefont {Kokail}}, \bibinfo {author} {\bibfnamefont {S.}~\bibnamefont {Flannigan}}, \bibinfo {author} {\bibfnamefont {N.}~\bibnamefont {Pearson}}, \bibinfo {author} {\bibfnamefont {M.}~\bibnamefont {Troyer}},\ and\ \bibinfo {author} {\bibfnamefont {P.}~\bibnamefont {Zoller}},\ }\bibfield  {title} {\bibinfo {title} {Practical quantum advantage in quantum simulation},\ }\href@noop {} {\bibfield  {journal} {\bibinfo  {journal} {Nature}\ }\textbf {\bibinfo {volume} {607}},\ \bibinfo {pages} {667} (\bibinfo {year} {2022})}\BibitemShut {NoStop}%
\bibitem [{\citenamefont {Whitfield}\ \emph {et~al.}(2011)\citenamefont {Whitfield}, \citenamefont {Biamonte},\ and\ \citenamefont {Aspuru-Guzik}}]{whitfield2011simulation}%
  \BibitemOpen
  \bibfield  {author} {\bibinfo {author} {\bibfnamefont {J.~D.}\ \bibnamefont {Whitfield}}, \bibinfo {author} {\bibfnamefont {J.}~\bibnamefont {Biamonte}},\ and\ \bibinfo {author} {\bibfnamefont {A.}~\bibnamefont {Aspuru-Guzik}},\ }\bibfield  {title} {\bibinfo {title} {Simulation of electronic structure hamiltonians using quantum computers},\ }\href@noop {} {\bibfield  {journal} {\bibinfo  {journal} {Molecular Physics}\ }\textbf {\bibinfo {volume} {109}},\ \bibinfo {pages} {735} (\bibinfo {year} {2011})}\BibitemShut {NoStop}%
\bibitem [{\citenamefont {Cervera-Lierta}(2018)}]{cervera2018exact}%
  \BibitemOpen
  \bibfield  {author} {\bibinfo {author} {\bibfnamefont {A.}~\bibnamefont {Cervera-Lierta}},\ }\bibfield  {title} {\bibinfo {title} {Exact ising model simulation on a quantum computer},\ }\href@noop {} {\bibfield  {journal} {\bibinfo  {journal} {Quantum}\ }\textbf {\bibinfo {volume} {2}},\ \bibinfo {pages} {114} (\bibinfo {year} {2018})}\BibitemShut {NoStop}%
\bibitem [{\citenamefont {Smith}\ \emph {et~al.}(2019)\citenamefont {Smith}, \citenamefont {Kim}, \citenamefont {Pollmann},\ and\ \citenamefont {Knolle}}]{smith2019simulating}%
  \BibitemOpen
  \bibfield  {author} {\bibinfo {author} {\bibfnamefont {A.}~\bibnamefont {Smith}}, \bibinfo {author} {\bibfnamefont {M.}~\bibnamefont {Kim}}, \bibinfo {author} {\bibfnamefont {F.}~\bibnamefont {Pollmann}},\ and\ \bibinfo {author} {\bibfnamefont {J.}~\bibnamefont {Knolle}},\ }\bibfield  {title} {\bibinfo {title} {Simulating quantum many-body dynamics on a current digital quantum computer},\ }\href@noop {} {\bibfield  {journal} {\bibinfo  {journal} {npj Quantum Information}\ }\textbf {\bibinfo {volume} {5}},\ \bibinfo {pages} {106} (\bibinfo {year} {2019})}\BibitemShut {NoStop}%
\bibitem [{\citenamefont {Brown}\ \emph {et~al.}(2010)\citenamefont {Brown}, \citenamefont {Munro},\ and\ \citenamefont {Kendon}}]{e12112268}%
  \BibitemOpen
  \bibfield  {author} {\bibinfo {author} {\bibfnamefont {K.~L.}\ \bibnamefont {Brown}}, \bibinfo {author} {\bibfnamefont {W.~J.}\ \bibnamefont {Munro}},\ and\ \bibinfo {author} {\bibfnamefont {V.~M.}\ \bibnamefont {Kendon}},\ }\bibfield  {title} {\bibinfo {title} {Using quantum computers for quantum simulation},\ }\href {https://doi.org/10.3390/e12112268} {\bibfield  {journal} {\bibinfo  {journal} {Entropy}\ }\textbf {\bibinfo {volume} {12}},\ \bibinfo {pages} {2268} (\bibinfo {year} {2010})}\BibitemShut {NoStop}%
\bibitem [{\citenamefont {Preskill}(2018)}]{Preskill2018quantumcomputingin}%
  \BibitemOpen
  \bibfield  {author} {\bibinfo {author} {\bibfnamefont {J.}~\bibnamefont {Preskill}},\ }\bibfield  {title} {\bibinfo {title} {Quantum {C}omputing in the {NISQ} era and beyond},\ }\href {https://doi.org/10.22331/q-2018-08-06-79} {\bibfield  {journal} {\bibinfo  {journal} {{Quantum}}\ }\textbf {\bibinfo {volume} {2}},\ \bibinfo {pages} {79} (\bibinfo {year} {2018})}\BibitemShut {NoStop}%
\bibitem [{\citenamefont {Kim}\ \emph {et~al.}(2023{\natexlab{b}})\citenamefont {Kim}, \citenamefont {Eddins}, \citenamefont {Anand}, \citenamefont {Wei}, \citenamefont {van~den Berg}, \citenamefont {Rosenblatt}, \citenamefont {Nayfeh}, \citenamefont {Wu}, \citenamefont {Zaletel}, \citenamefont {Temme},\ and\ \citenamefont {Kandala}}]{Kim2023}%
  \BibitemOpen
  \bibfield  {author} {\bibinfo {author} {\bibfnamefont {Y.}~\bibnamefont {Kim}}, \bibinfo {author} {\bibfnamefont {A.}~\bibnamefont {Eddins}}, \bibinfo {author} {\bibfnamefont {S.}~\bibnamefont {Anand}}, \bibinfo {author} {\bibfnamefont {K.~X.}\ \bibnamefont {Wei}}, \bibinfo {author} {\bibfnamefont {E.}~\bibnamefont {van~den Berg}}, \bibinfo {author} {\bibfnamefont {S.}~\bibnamefont {Rosenblatt}}, \bibinfo {author} {\bibfnamefont {H.}~\bibnamefont {Nayfeh}}, \bibinfo {author} {\bibfnamefont {Y.}~\bibnamefont {Wu}}, \bibinfo {author} {\bibfnamefont {M.}~\bibnamefont {Zaletel}}, \bibinfo {author} {\bibfnamefont {K.}~\bibnamefont {Temme}},\ and\ \bibinfo {author} {\bibfnamefont {A.}~\bibnamefont {Kandala}},\ }\bibfield  {title} {\bibinfo {title} {Evidence for the utility of quantum computing before fault tolerance},\ }\href {https://doi.org/10.1038/s41586-023-06096-3} {\bibfield  {journal} {\bibinfo  {journal} {Nature}\ }\textbf {\bibinfo {volume} {618}},\ \bibinfo {pages} {500} (\bibinfo {year}
  {2023}{\natexlab{b}})}\BibitemShut {NoStop}%
\bibitem [{\citenamefont {Roy}\ and\ \citenamefont {Harper}(2017)}]{PhysRevB.96.155118}%
  \BibitemOpen
  \bibfield  {author} {\bibinfo {author} {\bibfnamefont {R.}~\bibnamefont {Roy}}\ and\ \bibinfo {author} {\bibfnamefont {F.}~\bibnamefont {Harper}},\ }\bibfield  {title} {\bibinfo {title} {Periodic table for floquet topological insulators},\ }\href {https://doi.org/10.1103/PhysRevB.96.155118} {\bibfield  {journal} {\bibinfo  {journal} {Phys. Rev. B}\ }\textbf {\bibinfo {volume} {96}},\ \bibinfo {pages} {155118} (\bibinfo {year} {2017})}\BibitemShut {NoStop}%
\bibitem [{\citenamefont {Bell}\ and\ \citenamefont {Dean}(1970)}]{DF9705000055}%
  \BibitemOpen
  \bibfield  {author} {\bibinfo {author} {\bibfnamefont {R.~J.}\ \bibnamefont {Bell}}\ and\ \bibinfo {author} {\bibfnamefont {P.}~\bibnamefont {Dean}},\ }\bibfield  {title} {\bibinfo {title} {Atomic vibrations in vitreous silica},\ }\href {https://doi.org/10.1039/DF9705000055} {\bibfield  {journal} {\bibinfo  {journal} {Discuss. Faraday Soc.}\ }\textbf {\bibinfo {volume} {50}},\ \bibinfo {pages} {55} (\bibinfo {year} {1970})}\BibitemShut {NoStop}%
\bibitem [{\citenamefont {Wegner}(1980)}]{Wegner1980InversePR}%
  \BibitemOpen
  \bibfield  {author} {\bibinfo {author} {\bibfnamefont {F.~J.}\ \bibnamefont {Wegner}},\ }\bibfield  {title} {\bibinfo {title} {Inverse participation ratio in 2+$\epsilon$ dimensions},\ }\href {https://api.semanticscholar.org/CorpusID:120736426} {\bibfield  {journal} {\bibinfo  {journal} {Zeitschrift f{\"u}r Physik B Condensed Matter}\ }\textbf {\bibinfo {volume} {36}},\ \bibinfo {pages} {209} (\bibinfo {year} {1980})}\BibitemShut {NoStop}%
\bibitem [{\citenamefont {Abrahams}\ \emph {et~al.}(1979)\citenamefont {Abrahams}, \citenamefont {Anderson}, \citenamefont {Licciardello},\ and\ \citenamefont {Ramakrishnan}}]{PhysRevLett.42.673}%
  \BibitemOpen
  \bibfield  {author} {\bibinfo {author} {\bibfnamefont {E.}~\bibnamefont {Abrahams}}, \bibinfo {author} {\bibfnamefont {P.~W.}\ \bibnamefont {Anderson}}, \bibinfo {author} {\bibfnamefont {D.~C.}\ \bibnamefont {Licciardello}},\ and\ \bibinfo {author} {\bibfnamefont {T.~V.}\ \bibnamefont {Ramakrishnan}},\ }\bibfield  {title} {\bibinfo {title} {Scaling theory of localization: Absence of quantum diffusion in two dimensions},\ }\href {https://doi.org/10.1103/PhysRevLett.42.673} {\bibfield  {journal} {\bibinfo  {journal} {Phys. Rev. Lett.}\ }\textbf {\bibinfo {volume} {42}},\ \bibinfo {pages} {673} (\bibinfo {year} {1979})}\BibitemShut {NoStop}%
\bibitem [{\citenamefont {Kappus}\ and\ \citenamefont {Wegner}(1981)}]{kappus1981anomaly}%
  \BibitemOpen
  \bibfield  {author} {\bibinfo {author} {\bibfnamefont {M.}~\bibnamefont {Kappus}}\ and\ \bibinfo {author} {\bibfnamefont {F.}~\bibnamefont {Wegner}},\ }\bibfield  {title} {\bibinfo {title} {Anomaly in the band centre of the one-dimensional anderson model},\ }\href@noop {} {\bibfield  {journal} {\bibinfo  {journal} {Zeitschrift f{\"u}r Physik B Condensed Matter}\ }\textbf {\bibinfo {volume} {45}},\ \bibinfo {pages} {15} (\bibinfo {year} {1981})}\BibitemShut {NoStop}%
\bibitem [{\citenamefont {Stacey}(1982)}]{PhysRevD.26.468}%
  \BibitemOpen
  \bibfield  {author} {\bibinfo {author} {\bibfnamefont {R.}~\bibnamefont {Stacey}},\ }\bibfield  {title} {\bibinfo {title} {Eliminating lattice fermion doubling},\ }\href {https://doi.org/10.1103/PhysRevD.26.468} {\bibfield  {journal} {\bibinfo  {journal} {Phys. Rev. D}\ }\textbf {\bibinfo {volume} {26}},\ \bibinfo {pages} {468} (\bibinfo {year} {1982})}\BibitemShut {NoStop}%
\bibitem [{\citenamefont {Kogut}(1983)}]{RevModPhys.55.775}%
  \BibitemOpen
  \bibfield  {author} {\bibinfo {author} {\bibfnamefont {J.~B.}\ \bibnamefont {Kogut}},\ }\bibfield  {title} {\bibinfo {title} {The lattice gauge theory approach to quantum chromodynamics},\ }\href {https://doi.org/10.1103/RevModPhys.55.775} {\bibfield  {journal} {\bibinfo  {journal} {Rev. Mod. Phys.}\ }\textbf {\bibinfo {volume} {55}},\ \bibinfo {pages} {775} (\bibinfo {year} {1983})}\BibitemShut {NoStop}%
\bibitem [{\citenamefont {Nielsen}(2005)}]{Nielsen2005TheFC}%
  \BibitemOpen
  \bibfield  {author} {\bibinfo {author} {\bibfnamefont {M.~A.}\ \bibnamefont {Nielsen}},\ }\bibfield  {title} {\bibinfo {title} {The fermionic canonical commutation relations and the jordan-wigner transform}\ }(\bibinfo {year} {2005})\BibitemShut {NoStop}%
\bibitem [{\citenamefont {Nielsen}\ and\ \citenamefont {Chuang}(2010)}]{nielsen2010quantum}%
  \BibitemOpen
  \bibfield  {author} {\bibinfo {author} {\bibfnamefont {M.}~\bibnamefont {Nielsen}}\ and\ \bibinfo {author} {\bibfnamefont {I.}~\bibnamefont {Chuang}},\ }\href {https://books.google.co.kr/books?id=-s4DEy7o-a0C} {\emph {\bibinfo {title} {Quantum Computation and Quantum Information: 10th Anniversary Edition}}}\ (\bibinfo  {publisher} {Cambridge University Press},\ \bibinfo {year} {2010})\BibitemShut {NoStop}%
\bibitem [{\citenamefont {Suzuki}(1976)}]{suzuki1976generalized}%
  \BibitemOpen
  \bibfield  {author} {\bibinfo {author} {\bibfnamefont {M.}~\bibnamefont {Suzuki}},\ }\bibfield  {title} {\bibinfo {title} {Generalized trotter's formula and systematic approximants of exponential operators and inner derivations with applications to many-body problems},\ }\href@noop {} {\bibfield  {journal} {\bibinfo  {journal} {Communications in Mathematical Physics}\ }\textbf {\bibinfo {volume} {51}},\ \bibinfo {pages} {183} (\bibinfo {year} {1976})}\BibitemShut {NoStop}%
\bibitem [{\citenamefont {Hatano}\ and\ \citenamefont {Suzuki}(2005)}]{Hatano2005}%
  \BibitemOpen
  \bibfield  {author} {\bibinfo {author} {\bibfnamefont {N.}~\bibnamefont {Hatano}}\ and\ \bibinfo {author} {\bibfnamefont {M.}~\bibnamefont {Suzuki}},\ }\bibinfo {title} {Finding exponential product formulas of higher orders},\ in\ \href {https://doi.org/10.1007/11526216_2} {\emph {\bibinfo {booktitle} {Quantum Annealing and Other Optimization Methods}}},\ \bibinfo {editor} {edited by\ \bibinfo {editor} {\bibfnamefont {A.}~\bibnamefont {Das}}\ and\ \bibinfo {editor} {\bibfnamefont {B.}~\bibnamefont {K.~Chakrabarti}}}\ (\bibinfo  {publisher} {Springer Berlin Heidelberg},\ \bibinfo {address} {Berlin, Heidelberg},\ \bibinfo {year} {2005})\ pp.\ \bibinfo {pages} {37--68}\BibitemShut {NoStop}%
\end{thebibliography}%

\section{Acknowledgement} 
This research was supported by the quantum computing technology development program of the National Research Foundation of Korea(NRF) funded by the Korean government (Ministry of Science and ICT(MSIT)) (No. 2020M3H3A1110365).
This work was supported by the National Research Foundation of Korea
(NRF) grant funded by the Korea government (MSIT) (Grants No. RS-2023-00252085 and No. RS-2023-00218998).

\pagebreak
\newpage
\renewcommand{\thefigure}{S\arabic{figure}}
\setcounter{figure}{0}
\renewcommand{\theequation}{S\arabic{equation}}
\setcounter{equation}{0}

\begin{widetext}
\section{Supplementary material for\\``Floquet Chiral Quantum Walk in Quantum Computer"}

\maketitle

\section{Details on quantum circuit transcription}
\textit{Quantum device hardware}- We utilize IBM superconducting transmon qubit quantum processors. OpenQASM3 27-qubit device \textit{ibm\_hanoi}(QV=64) is used for simulating the discrete time evolution of the FCQW model. OpenQASM3 27-qubit device \textit{ibmq\_kolkata}(QV=128) is used for the time evolution of the non-chiral model. Here, quantum Volume (QV) indicates a metric that quantifies the largest size of a quantum circuit that can be effectively executed on a given quantum computer. It serves as an indicator of the system's computational capabilities and error rates. 

Detailed layout of the qubit topology and calibration data of the quantum devices used in our experiments are provided in Fig. \ref{figs1}. In addition, we have utilized the dynamical decoupling technique to reduce the quantum error. We utilize the Qiskit transpiler to optimize our circuit to real(fake) backend environments. We execute a total 7000 shots for each implementation. Chiral simulation is executed on Oct 3, 2023, 08:44 - 10:21(GMT) by \textit{ibm\_hanoi}. The non-Chiral case is also simulated on Oct 6, 2023, 11:33 - 13:41(GMT) by \textit{ibmq\_kolkata}.

\textit{Noise simulation}-
To compare with the real device simulation, we conducted the model simulation using Qiskit's built-in noise simulation with the fake provider and fake backends. Fake backends mimic the behavior of real IBM quantum computers and they contain important information about real quantum devices(basis gate, circuit topology, gate-type-dependent intrinsic error rates, etc.) for testing simulation performance in noisy environments. 

\begin{figure*}[ht]
\centering
\includegraphics[width=\linewidth]{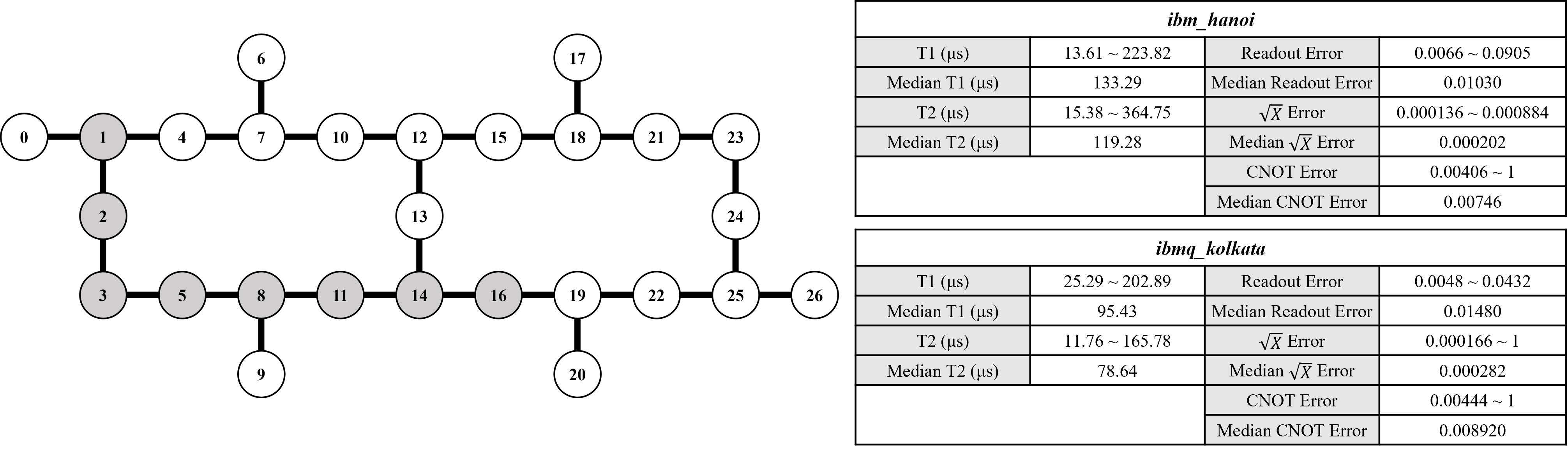}
\caption{The illustration of qubit topology of 27-qubit quantum devices and calibration information of \textit{ibm\_hanoi} and \textit{ibmq\_kolkata}. The qubits which are highlighted in a gray color show an example of an 8-qubit chain used for $L=8$ time evolution simulations. T1 and T2 are the relaxation and dephasing times respectively of the qubits. T1, T2, readout errors, and $\sqrt{X}$ errors are single-qubit calibration and CNOT error is about qubit-pairs.}
    \label{figs1}
\end{figure*}
\section{Implementation of non-chiral Hamiltonian}

\textit{Jordan-Wigner Transformation}--
We aim to simulate the one-dimensional tight-binding Hamiltonian, which is given as, 
\begin{gather}\label{hamiltonian1}
{
\hat{H} = \sum_{i} -[\hat{c}_{i}^{\dagger} \hat{c}_{i+1} + \hat{c}_{i+1}^{\dagger} \hat{c}_{i}] + Wu_{i} \hat{c}_{i}^{\dagger} \hat{c}_{i}
}
\end{gather}
where $\hat{c}_{i}^{\dagger}$ and $\hat{c}_{i}$ are creation and annihilation operator of the fermion. $W$ is onsite-potential strength. $u_{i}$ determines the distribution of on-site potentials. To implement the time-evolution operator $\hat{U}=e^{i\hat{H}t}$, we transcribe $\hat{H}$ and $\hat{U}$ to Pauli matrices representation by Jordan-Wigner transformation that is described below : 
\begin{gather}
{
\hat{c}_j = -\left( \bigotimes_{k=1}^{j-1}\hat{\sigma}_{k}^{z} \right)\otimes \tilde{\sigma}_j
}
\end{gather}
where $\hat{\sigma}^{\mu}_{j}$ is $\mu$-th Pauli operator acting on the $j$-th qubit. $\tilde{\sigma}_j$ is used to denote the qubit operator $|0\rangle \langle1|$ acting on the $j$-th qubit\cite{Nielsen2005TheFC}. With the result of simple algebra, we can show each term of Hamiltonian (Eq.\eqref{hamiltonian1}) can be transformed as,
\begin{gather}
c_j^{\dagger} c_{j+1} + c_{j+1}^{\dagger} c_j \longrightarrow \hat{\sigma}_{j}^{X} \hat{\sigma}_{j+1}^{X} + \hat{\sigma}_{j}^{y} \hat{\sigma}_{j+1}^{y} ,
\quad 
c_{j}^{\dagger} c_j \longrightarrow \hat{\sigma}_{j}^{z}
\end{gather}
respectively. As a result, the tight-binding Hamiltonian in Eq.\eqref{hamiltonian1} can be re-written in terms of the spin model as,
\begin{gather}
{
\hat{H} = \sum_{i} -J[\hat{\sigma}_{i}^{x} \hat{\sigma}_{i+1}^{x} + \hat{\sigma}_{i}^{y} \hat{\sigma}_{i+1}^{y}] + Wu_{i} \hat{\sigma}_{i}^{z}
}
\end{gather}

\textit{Quantum circuit representation}--
To implement the time evolution operator $\hat{U}=e^{-i\hat{H}t}$, we construct a trotterized circuit composed of the hopping and the potential terms separately. In quantum circuit language, the unitary time evolution operator of the hopping term of Hamiltonian can be constructed by the successive applications of the two CNOT gates and the one RZ gate as\cite{nielsen2010quantum},
\begin{equation} 
{
e^{-i(-J \hat{\sigma}_{i}^{x} \hat{\sigma}_{i+1}^{x})dt} \Longrightarrow 
\begin{tabular}{r}
\vspace{.1em}\\
\Qcircuit @C=0.9em @R=1em {
& & & &\lstick{\ket{i}} &  \qw & \gate{H} & \ctrl{1} & \qw & \ctrl{1} & \gate{H} &\qw \\ 
& & & &\lstick{\ket{i+1}} & \qw & \gate{H} &\targ & \gate{R_z(-Jdt)} & \targ& \gate{H} &\qw
}
\vspace{1.5em}\hspace{1em}\\
\end{tabular}
}
\end{equation}
In addition, the onsite potential term of Hamiltonian can be represented as,
\begin{equation}
{
e^{-i(-J \hat{\sigma}_{i}^{y} \hat{\sigma}_{i+1}^{y})dt} \Longrightarrow 
\begin{tabular}{r}
\vspace{.1em}\\
\Qcircuit @C=0.9em @R=1em {
&&& &\lstick{\ket{i}} &  \qw & \gate{H_y} & \ctrl{1} & \qw & \ctrl{1} & \gate{H_y} &\qw \\ 
&&& &\lstick{\ket{i+1}} & \qw & \gate{H_y} &\targ & \gate{R_z(-Jdt)} & \targ& \gate{H_y} &\qw
}
\vspace{1.5em}\hspace{1em}\\
\end{tabular}
}
\end{equation}
where $H_{y}$ is Hadamard-$y$ single-qubit rotating gate which can be used to perform a change $y$-basis to $z$-basis, satisfying $H_{y}ZH_{y}=Y$. Finally, the qubit representation of the last term can be constructed with the single RZ gate.

\textit{Trotterization}-
For generic operator, $\hat{A}$ and $\hat{B}$, $e^{i(\hat{A}+\hat{B})t}$ is not equivalent to $e^{i\hat{A}t}e^{i\hat{B}t}$ except when the two operators coummute as $[\hat{A},\hat{B}]=0$. However, if Hamiltonian is given by sum of $m$ different operators such that $\hat{H}=\sum_{i=1}^{m}\hat{H}_{i}$, the unitary time-evolution operator  $\hat{U}=e^{-i\hat{H}t}$ can be approximated using a Trotter-Suzuki formula. Trotter-Suzuki decomposition is given by :
\begin{gather}
{
e^{(\hat{A}+\hat{B})t} = e^{\hat{A}t}e^{\hat{B}t}+O(t^2)
}.
\end{gather}
We can separate the time evolution operator with first-order Trotter-Suzuki approximation as,
\begin{gather}
{
e^{-i\hat{H}t} \sim \prod_{i=1}^{m} \left( e^{-i\hat{H_i}t/n}\right)^n}
\end{gather}
where n is trotter order and error is proportional to $O(1/n^2)$\cite{suzuki1976generalized,Hatano2005}. We decompose our time evolution operator to $3N\times n$ terms, $U_x^i = e^{-i(-J \hat{\sigma}_{i}^{x} \hat{\sigma}_{i+1}^{x})t/n}$, $U_y^i = e^{-i(-J \hat{\sigma}_{i}^{y} \hat{\sigma}_{i+1}^{y})t/n}$, and  $U_z^i = e^{-i(Wu_i \hat{\sigma}_{i}^{z})t/n}$(see Fig. \ref{figs2}). In the case of large $n$, trotter error can be negligible, however, the intrinsic error rate of quantum devices is rapidly escalated by an increment of non-local gates. In such case, Device error is more dominant than trotter error, we simulate the $n=2$ case for our experiments. The initial state was chosen such that the particle is positioned at site 4, and we measure the time evolution in six intervals from $t=0.1$ to $t=2.0$.
\begin{figure*}[ht]
\centering
\includegraphics[width=\linewidth]{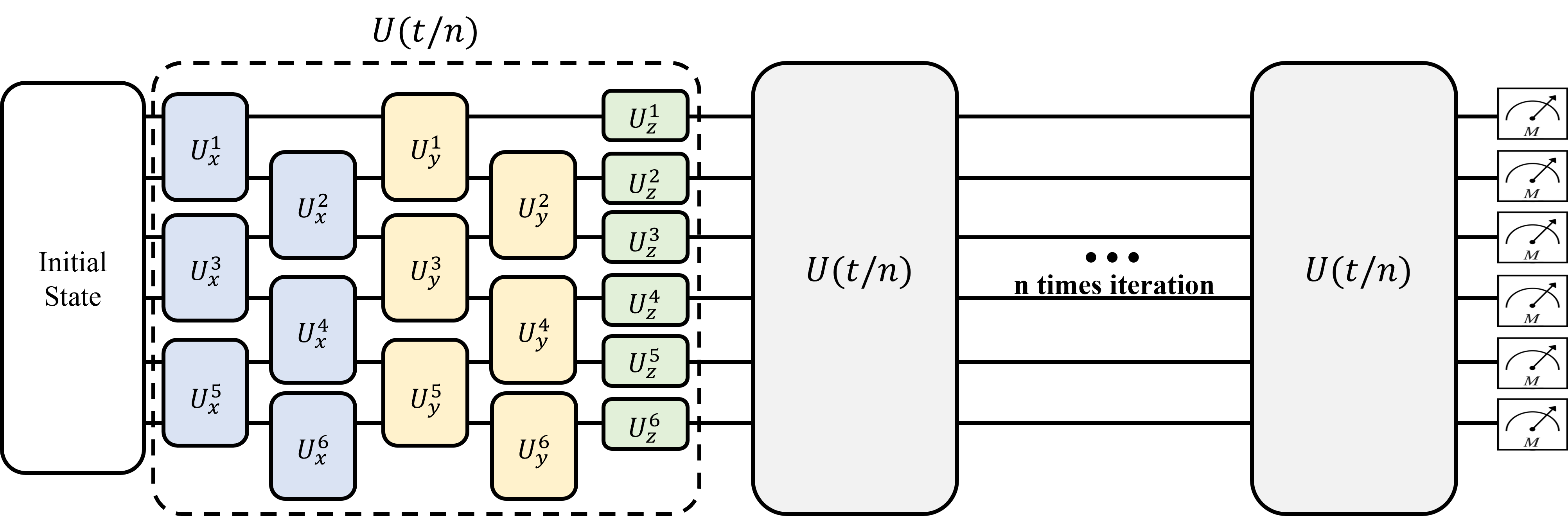}
\caption{The illustrative figure of a non-chiral lattice model of trotter circuit}
    \label{figs2}
\end{figure*}

\end{widetext}

\end{document}